\documentclass{emulateapj}

\begin{document}

\shortauthors{Luhman}
\shorttitle{Binary Brown Dwarf at 2 pc}

\title{Discovery of a Binary Brown Dwarf at 2 Parsecs from the Sun\altaffilmark{1}}

\author{
K. L. Luhman\altaffilmark{2,3}}

\altaffiltext{1}
{Based on data from the {\it Wide-field Infrared Survey Explorer},
Gemini Observatory, the Two Micron All-Sky Survey, the Deep Near-Infrared
Survey of the Southern Sky, and the Digitized Sky Survey.}

\altaffiltext{2}{Department of Astronomy and Astrophysics,
The Pennsylvania State University, University Park, PA 16802, USA;
kluhman@astro.psu.edu}
\altaffiltext{3}{Center for Exoplanets and Habitable Worlds, The 
Pennsylvania State University, University Park, PA 16802, USA}

\begin{abstract}

I am using multi-epoch astrometry from the {\it Wide-field Infrared Survey
Explorer (WISE)} to search for new members of the solar neighborhood via
their high proper motions. 
Through this work, I have identified WISE J104915.57-531906.1 as a 
high proper motion object and have found additional detections in
images from the Digitized Sky Survey, the Two Micron All-Sky Survey,
and the Deep Near-Infrared Survey of the Southern Sky.
I have measured a parallax of 0.496$\pm0.037\arcsec$ (2.0$\pm$0.15~pc)
from the astrometry in these surveys, making WISE J104915.57-531906.1
the third closest system to the Sun. During spectroscopic observations
with GMOS at Gemini Observatory, an $i$-band acquisition image
resolved it as a $1\farcs5$ (3~AU) binary.
A spectrum was collected for the primary, which I classify as L8$\pm$1.
The secondary is probably near the L/T transition as well given that
it is only modestly fainter than the primary ($\Delta i=0.45$~mag).

\end{abstract}


\section{Introduction}

Stars in the solar neighborhood have been identified primarily through their
large proper motions
\citep[e.g.,][]{bar16,wol19,ros26,van44,gic71,luy79,lep05b}.
Most brown dwarfs, on the other hand, have been found through their
distinctive colors \citep[e.g.,][]{bur04,cru07,burn10,alb11,kir11}
because wide-field infrared (IR) surveys initially offered limited
data on proper motions.
However, a significant amount of multi-epoch astrometry is now available
from various combinations of the Two Micron All-Sky Survey
\citep[2MASS,][]{skr06}, the Deep Near-Infrared Survey of the Southern
Sky \citep[DENIS,][]{ep99}, the Sloan Digital Sky Survey \cite[SDSS,][]{yor00},
the United Kingdom Infrared Telescope Infrared Deep Sky Survey
\citep[UKIDSS,][]{law07}, Pan-STARRS1 \citep[PS1,][]{kai02}, and the
survey performed by the {\it Wide-field Infrared Survey Explorer}
\citep[{\it WISE},][]{wri10}.  As a result, the role of proper motions
in brown dwarf searches is expanding rapidly
\citep{she09,art10,kir10,dea09,dea11,liu11,giz11,sch11,sch12}.

The {\it WISE} satellite imaged each position in the sky at multiple epochs
that span 6--12 months, which enables proper motion measurements with
{\it WISE} data alone.
The accuracy of {\it WISE} astrometry (0\farcs2 at SNR$>$50) is lower than
that of other modern wide-field surveys, but it provides the best sensitivity
to the coolest brown dwarfs \citep{cus11}.
In addition, it is easier to detect the highest proper motions exhibited by
stars ($\mu\sim1$--$10\arcsec$~yr$^{-1}$) through multiple identical mappings
across a baseline of less than a year
than with a comparison of surveys conducted at different wavelengths
and at more widely separated epochs.
Through an analysis of the multi-epoch astrometry from {\it WISE}, I have
discovered one the closest neighbors of the Sun, which I present in this
Letter.

\section{Proper Motion Survey with {\it WISE}}

{\it WISE} was designed to complete a map of the entire sky in 6 months.
Given that its mission lifetime was nearly 13 months, it was able to
perform two full surveys of the sky and begin a third one.
In each 6 month survey, {\it WISE} obtained one 8.8~sec exposure
in each operable band every 1.5~hours for a period ranging from 12~hours
near the ecliptic plane to 6 months at the ecliptic poles.
The four bands of {\it WISE} were centered at 
3.4, 4.6, 12, and 22~\micron, which are denoted as $W1$, $W2$, $W3$, and $W4$,
respectively.
All of the bands were employed from 2010 January 7 through 2010 August 6.
Because of the successive exhaustion of the outer and inner cryogen
tanks, images were not collected through $W4$ and $W3$ after 2010 August 6
and 2010 September 29, respectively. The satellite continued to observe
in $W1$ and $W2$ until 2011 February 1 \citep{mai11a}.

To search for high proper motion objects with {\it WISE},
I began by retrieving astrometry and photometry
from the All-Sky Single Exposure (L1b) Source Table, the 
3-Band Cryo Single Exposure (L1b) Source Table, and the Preliminary
Post-Cryo Single Exposure (L1b) Source Table for all sources that lacked
2MASS counterparts within $3\arcsec$. I chose to omit sources with
close 2MASS counterparts to focus my search on objects with either high
proper motions ($\mu\gtrsim0\farcs3$~yr$^{-1}$) or low temperatures
($\gtrsim$T7 at 10~pc). Using TOPCAT\footnote{http://www.starlink.ac.uk/topcat/}
and STILTS\footnote{http://www.starlink.ac.uk/stilts/} \citep{tay05,tay06},
I identified all groups of detections that had separations of $<1\farcs5$
and $<1$~day between neighboring detections. Thus, these groups were
designed to contain all detections of a source during the time that
it was observed continuously. Each of these periods is referred to as
an ``epoch" in the remaining discussion.

\begin{figure}
\epsscale{1}
\plotone{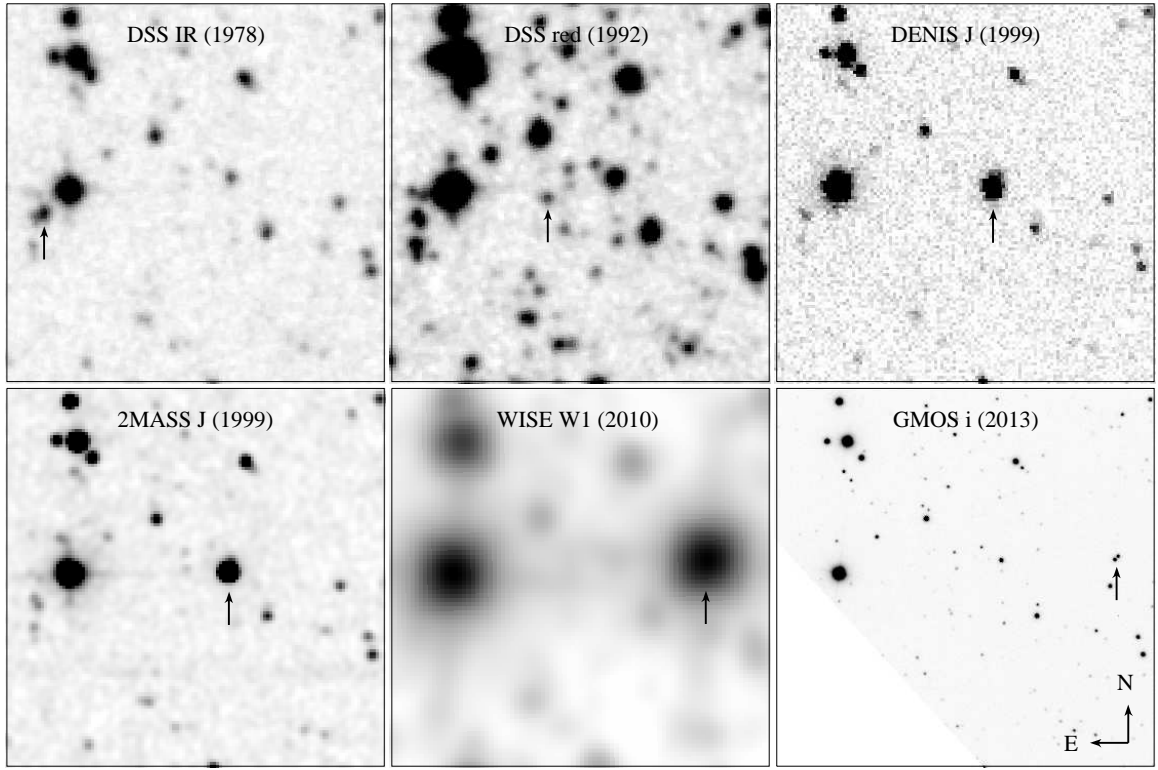}
\caption{
Detections of WISE~1049-5319 in images from DSS, 2MASS, DENIS, {\it WISE},
and GMOS (arrows). It is resolved as a $1\farcs5$ binary by GMOS.
The size of each image is $2\arcmin\times2\arcmin$.
}
\label{fig:image}
\end{figure}

Using the average coordinates of detections within a given epoch, 
I matched sources from different epochs, computed the proper motions
($\mu$) for these matches, and identified the matches with $\mu$/$\sigma_\mu>3$.
Among the resulting high proper motion candidates, I omitted those
that have been found in previous studies and that appeared extended or
blended with other sources upon visual inspection.
Nearly all of the remaining candidates that were detected
in $W3$ and $W4$ exhibited very red colors that were indicative of galaxies
($W2-W3>2$, $W3-W4>1.5$),
and thus their putative motions were likely caused by resolved emission.
However, one of these candidates was only modestly red, and hence
closer to the colors observed for stellar sources.
This candidate corresponds to WISE~J104915.57$-$531906.1 (hereafter
WISE 1049$-$5319) and consisted of 12, 19, and 18 detections near
2010 January 9, 2010 July 2, and 2011 January 6, respectively.
I inspected images from 2MASS, DENIS, and the Digitized Sky Survey (DSS)
at the positions expected for WISE 1049$-$5319 given the proper motion
measured from the three {\it WISE} epochs. It is detected at
$J$, $H$, and $K_s$ (1.25, 1.65, 2.16~\micron) with 2MASS and at
$i$ (0.8~\micron), $J$, and $K_s$ with DENIS.
It appears in the red and IR (0.66, 0.85~\micron) images of DSS,
but not the blue image (0.45~\micron).
The detections of WISE 1049$-$5319 in DSS IR and red, 2MASS $J$, DENIS $J$, 
and {\it WISE} $W1$ are shown in Figure~\ref{fig:image}. I also include
the $i$-band acquisition image obtained during
spectroscopic observations of WISE 1049$-$5319 (Section~\ref{sec:phot}).
The {\it WISE} image is displayed on a logarithmic scale so that sources
near the detection limit can be seen while keeping WISE 1049$-$5319
relatively unsaturated.

\section{Characterization of WISE 1049$-$5319}

\subsection{Photometric and Spectroscopic Properties}
\label{sec:phot}

\begin{figure}
\epsscale{1}
\plotone{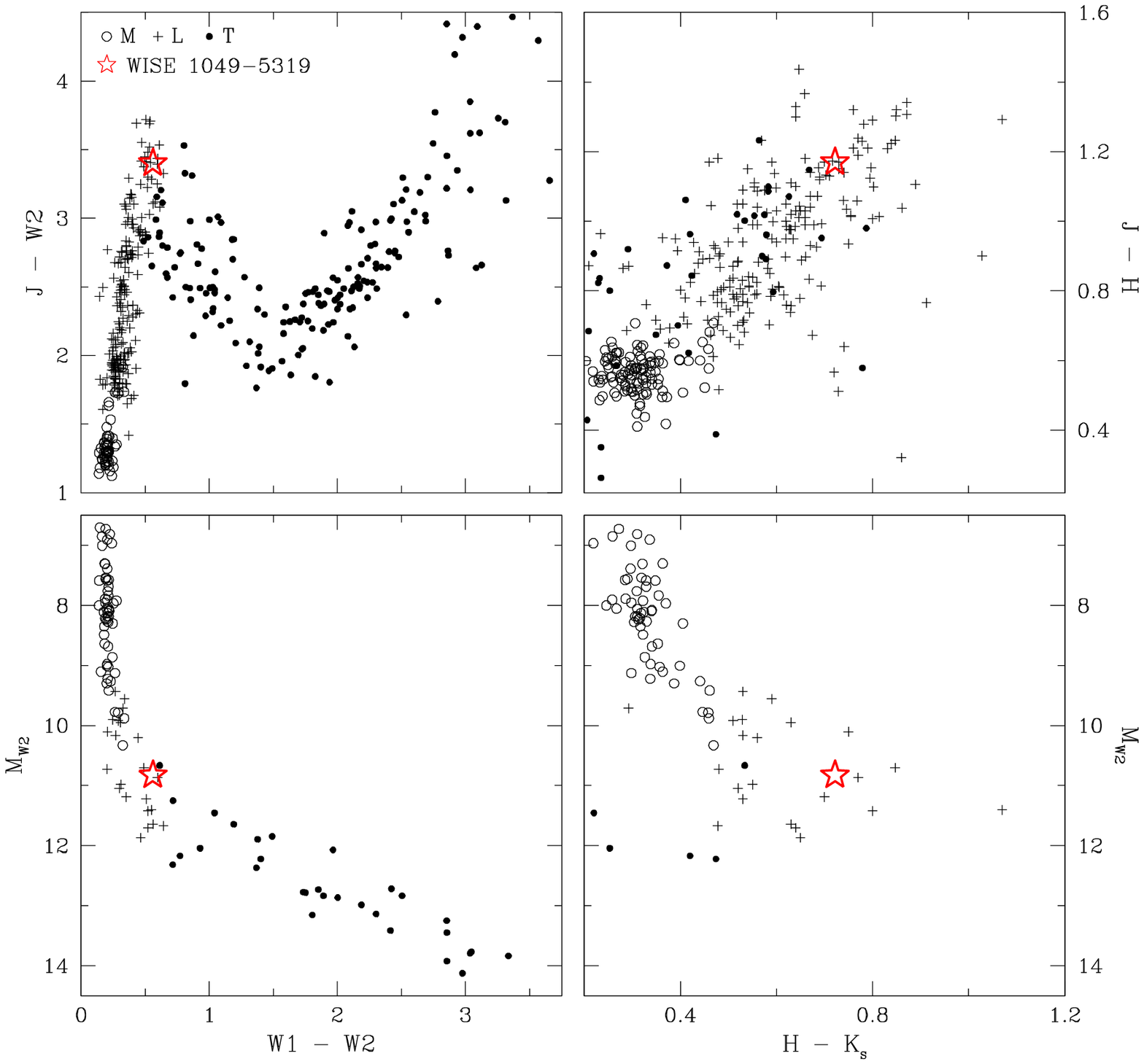}
\caption{
Color-magnitude and color-color diagrams for WISE~1049-5319
(star) and a sample of M, L, and T dwarfs \citep[open circles, crosses, and
filled circles,][http://DwarfArchives.org]{leg10a,kir11,kir12}.
}
\label{fig:cmd}
\end{figure}

To investigate the photometric properties of WISE 1049$-$5319, I have plotted
it on color-magnitude and color-color diagrams in Figure~\ref{fig:cmd} that
are constructed from 2MASS and {\it WISE} bands. The absolute magnitude
in $W2$ is computed with the parallax measured in Section~\ref{sec:astro}.
For comparison, I have included a sample of known M, L, and T dwarfs
\citep[][http://DwarfArchives.org]{leg10a,kir11,kir12}.
2MASS measurements are shown for these field dwarfs since the near-IR data
for WISE 1049$-$5319 are also from 2MASS.
All of the data for WISE 1049$-$5319 in Figure~\ref{fig:cmd}
are consistent with a dwarf between spectral types of $\sim$L5 and L9.
The position of WISE 1049$-$5319 in $M_{W2}$ versus $W1-W2$ and
is modestly brighter than the lower envelope of the L dwarf sequence,
which is suggestive of an unresolved binary. A similar offset is present
in various other IR color-magnitude diagrams as well.
Finally, I note that the entry for WISE 1049$-$5319 in the All-Sky Source
Catalog, which consists of the 31 detections from the first two epochs, is
flagged as a possible variable in $W1$ and $W2$. Thus, it may be an attractive
target for detailed photometric monitoring \citep[e.g.,][]{art09,rad12}.

\begin{figure}
\epsscale{1}
\plotone{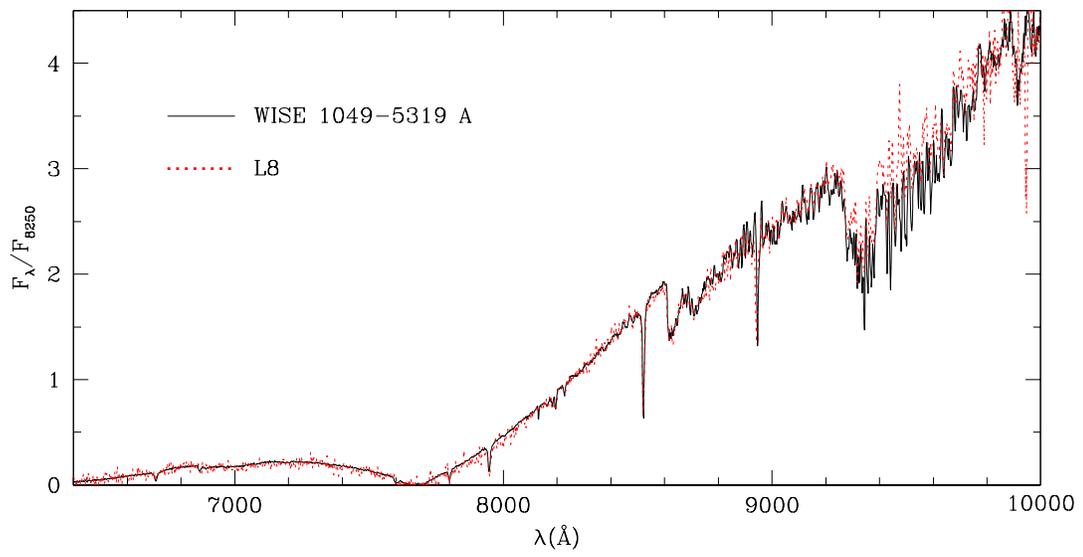}
\caption{
Optical spectrum of WISE~1049-5319~A compared to the L8 standard
2MASS J16322911+1904407 \citep{kir99}.
}
\label{fig:spec}
\end{figure}

To measure the spectral type of WISE 1049$-$5319, I pursued 
spectroscopy of it with the Gemini Multi-Object Spectrograph (GMOS)
at the Gemini South telescope on the night of 2013 February 23.
The observations began with a 10~s $i$-band exposure for target acquisition.
The FWHM of point sources in this image was $0\farcs6$.
At the position expected for WISE 1049$-$5319 given its proper motion,
a $1\farcs5$ pair with $\Delta i=0.45$~mag was detected.
I conclude that both objects are counterparts to WISE 1049$-$5319
and that they comprise a binary system since the images from DSS, DENIS,
and 2MASS do not detect a source at that location. The binarity of
WISE 1049$-$5319 is consistent with its overluminous position in
Figure~\ref{fig:cmd}.  Spectroscopy was performed on the brighter (southeast)
component, which I refer to as WISE 1049$-$5319~A.
The spectrograph was configured with the 400~l/mm grating, the RG610 order
blocking filter, and the $0\farcs75$ slit, which produced data with
a resolution of 5~\AA. The slit was aligned to the parallactic angle and
two 5~min exposures were collected. The resulting spectrum is shown in
Figure~\ref{fig:spec}. It matches closely with the spectrum of
the L8 dwarf standard 2MASS J16322911+1904407 from \citet{kir99}.
Based on a comparison to the other standards from that study, the
uncertainty in this classification is $\pm1$~subclass.
The hydrogen burning limit is predicted to occur at
$T_{\rm eff}\gtrsim1700$~K \citep{cha00,bur01} for ages less than 10~Gyr,
which corresponds to spectral types of $\lesssim$L5 \citep{gol04}.
Thus, WISE 1049$-$5319~A is very likely a brown dwarf. 
The presence of strong lithium absorption ($W_{\lambda}=8\pm1$~\AA)
provides further support for its substellar nature \citep{reb92}.
Given the flux ratio of the pair, the secondary is probably a brown dwarf
as well with a spectral type of late L or early T. 

\subsection{Parallax and Proper Motion}
\label{sec:astro}

The binarity of WISE 1049$-$5319 has implications for the measurement
of its parallax and proper motion.
In an unresolved binary containing unequal components, the orbital motion
is reflected in movement of the photocenter, which can introduce errors into
derived parallax and proper motion \citep[e.g.,][]{dup12}.
In the case of WISE 1049$-$5319,
its separation and the substellar masses of its components are indicative of
an orbital period of $\sim$25~years. Since this is much longer than the
baseline of one year across which parallactic motion occurs, the
measurement of parallax should be unaffected. Because the astrometry for
WISE 1049$-$5319 spans a timescale of decades, the measured proper motion
is more susceptible to an error from the motion of the photocenter, but
the effect is still quite small. For instance, using the $i$-band flux
ratio, the amplitude of the shift of the photocenter is only $\sim0.5$\% of
the total motion of WISE 1049$-$5319 over the period for which astrometry
is available. However, if the flux ratio of a system varies with wavelength,
then the position of the photocenter will do so as well, which can lead
to errors in both the proper motion and parallax if they are based on
multi-wavelength astrometry. To mitigate such errors, I have
excluded the GMOS astrometry from the measurement of
the parallax and proper motion. The other astrometric data are either from
IR bands ({\it WISE}, DENIS, 2MASS), within which the flux ratio
should not vary as much as in the optical, or already have uncertainties 
that are larger than this effect. 
Focusing on the IR astrometry offers an additional advantage; 
because optical-to-IR colors of L and T dwarfs become redder with
later types, the flux ratio of WISE 1049$-$5319 is probably closer
to unity at IR wavelengths than that measured with GMOS in the $i$ band,
corresponding to a smaller motion of the photocenter.

I adopted the astrometry from the 2MASS Point Source Catalog and the
Third DENIS Release. For each {\it WISE} epoch, I computed the mean of the
coordinates that were retrieved from the single-exposure catalogs,
with the exception of one detection from the second epoch with discrepant
coordinates. For the DSS images, I measured the positions of the counterpart
to WISE 1049$-$5319 with the IRAF task {\it starfind}.
The counterparts in 2MASS and DENIS are partially blended with a
star at a distance of $\sim3\arcsec$ to the north.
The latter should be $\sim5$~mag fainter than WISE 1049$-$5319 in the near-IR
bands of 2MASS and DENIS based on its optical photometry from the USNO-B1.0
Catalog \citep{mon03}. Thus, the astrometry from the 2MASS and DENIS should be
negligibly affected by the fainter star.
To place all astrometry on the same reference system, I computed the
average offset in right ascension and declination between 2MASS and
the other surveys for sources within a few arcminutes of WISE 1049$-$5319.
These offsets were then applied to the coordinates for WISE 1049$-$5319
from DSS, DENIS, and each of the {\it WISE} epochs to align them to the
2MASS system.

For the {\it WISE} astrometric errors, I adopted the standard deviations
of the right ascensions and declinations of the individual detections
at a given epoch. For 2MASS, I adopted the errors from the Point
Source Catalog. To estimate the errors for DSS and DENIS,
I assumed that their sum in quadrature with the errors from the
2MASS equaled the standard deviations of the differences in right
ascension and declination between DSS/2MASS and DENIS/2MASS
for stars near the flux of WISE 1049$-$5319. The detection in DSS IR is
partially blended with a fainter source to the east. In an attempt to account
for the effect on the astrometry, I doubled the error for its astrometry.
The astrometry for WISE 1049$-$5319 from DSS, 2MASS, DENIS, and {\it WISE}
is compiled in Table~\ref{tab:astro}.

\begin{figure}
\epsscale{1.2}
\plotone{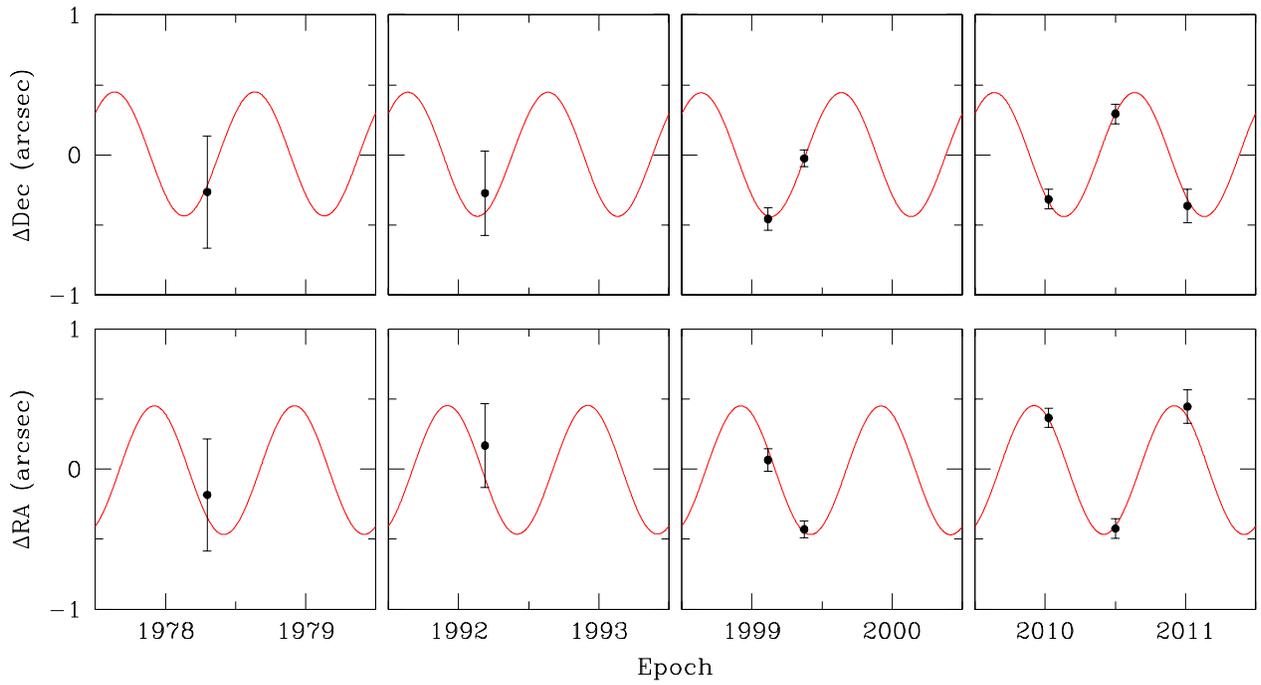}
\caption{
Relative astrometry of WISE~1049-5319 in images from DSS, DENIS, 2MASS, and
{\it WISE} (points) compared to the best-fit model of parallactic motion
(Table~\ref{tab:data}, red curve). The proper motion produced by the fitting
has been subtracted.
}
\label{fig:pm}
\end{figure}

I have performed least-squares fitting of proper and parallactic motion to the
astrometry for WISE 1049$-$5319 using the IDL program MPFIT.
The resulting measurements of the parallax and proper motion are presented
in Table~\ref{tab:data}. The parallax corresponds to a distance of
2.0$\pm$0.15~pc. To test the validity of the errors, I fitted 1000 sets of
astrometry that were generated by adding Gaussian noise to the measured
positions. The standard deviations of the resulting values of parallax and
proper motion agreed with the uncertainties produced by MPFIT.
In Figure~\ref{fig:pm}, the measured astrometric offsets after removal
of proper motion are plotted with the offsets are expected for the
derived value of the parallax.
Although the DSS data have large uncertainties, they do provide tight
constraints on the proper motion because of the long baseline, which in
turn allows the 2MASS and {\it WISE} data to better constrain the parallax.

\section{Discussion}

With a distance of 2.0$\pm$0.15~pc, WISE~1049$-$5319 is
the closest neighbor of the Sun that has been found in
nearly a century \citep{hen39,bar16,ada17,vou17}.
It is only slightly more distant than Barnard's star, which is the second
nearest known system \citep[1.834$\pm$0.001~pc,][]{ben99}.
The low galactic latitude of WISE~1049$-$5319 ($b=5\arcdeg$) is likely
the reason why it was not found in previous surveys for nearby brown dwarfs,
which have tended to avoid the galactic plane.
Because of its proximity to the Sun, WISE~1049$-$5319 is a unique target
for a variety of studies, such as direct imaging and radial velocity
searches for planets.

\acknowledgements
I acknowledge support from grant NNX12AI47G from the NASA Astrophysics
Data Analysis Program.
{\it WISE} is a joint project of the University of California, Los Angeles,
and the Jet Propulsion Laboratory (JPL)/California Institute of
Technology (Caltech), funded by NASA. 
The Gemini data were obtained
through program GN-2013A-DD-2. Gemini Observatory is operated by the 
Association of Universities for Research in Astronomy, Inc., under a
cooperative agreement with the NSF on behalf of the Gemini partnership:
the National Science Foundation (United States), the National Research Council
(Canada), CONICYT (Chile), the Australian Research Council (Australia),
Minist\'{e}rio da Ci\^{e}ncia, Tecnologia e Inova\c{c}\~{a}o (Brazil) and
Ministerio de Ciencia, Tecnolog\'{i}a e Innovaci\'{o}n Productiva (Argentina).
2MASS is a joint project of the University of
Massachusetts and the Infrared Processing and Analysis Center (IPAC) at
Caltech, funded by NASA and the NSF.
The Digitized Sky Survey was produced at the Space Telescope Science
Institute under U.S. Government grant NAG W-2166. The images of these
surveys are based on photographic data obtained using the Oschin Schmidt
Telescope on Palomar Mountain and the UK Schmidt Telescope. The plates
were processed into the present compressed digital form with the
permission of these institutions. 
This work uses data from 
the M, L, and T dwarf compendium at http://DwarfArchives.org
(maintained by Chris Gelino, Davy Kirkpatrick, and Adam Burgasser)
and the NASA/IPAC Infrared Science Archive (operated by JPL
under contract with NASA).
The Center for Exoplanets and Habitable Worlds is supported by the
Pennsylvania State University, the Eberly College of Science, and the
Pennsylvania Space Grant Consortium.
The DENIS project was partly funded by the SCIENCE and the HCM plans of
the European Commission under grants CT920791 and CT940627.
It was supported by INSU, MEN and CNRS in France, by the State of
Baden-W\"urttemberg in Germany, by DGICYT in Spain, by CNR in Italy, by
FFwFBWF in Austria, by FAPESP in Brazil, by OTKA grants F-4239 and F-013990
in Hungary, and by the ESO C\&EE grant A-04-046.
Jean Claude Renault from IAP was the Project manager. Observations were
carried out thanks to the contribution of numerous students and young
scientists from all involved institutes, under the supervision of P. Fouqu\'e.

\clearpage

\begin{deluxetable}{lllll}
\tabletypesize{\scriptsize}
\tablewidth{0pt}
\tablecaption{Astrometry for WISE~J104915.57$-$531906.1\label{tab:astro}}
\tablehead{
\colhead{$\alpha$ (J2000)} & \colhead{$\delta$ (J2000)} & \colhead{$\sigma_{\alpha,\delta}$} & \colhead{MJD} &  \colhead{Source} \\
\colhead{($\arcdeg$)} & \colhead{($\arcdeg$)} & \colhead{($\arcsec$)} & \colhead{} & \colhead{}}
\startdata
162.355972 & $-$53.321606 & 0.40 & 43617.5 & DSS IR\\
162.338312 & $-$53.320242 & 0.30 & 48690.6 & DSS red\\
162.329376 & $-$53.319612 & 0.08 & 51220.8 & DENIS\\
162.328814 & $-$53.319466 & 0.06 & 51315.1 & 2MASS\\
162.315519 & $-$53.318500 & 0.07 & 55205.3 & {\it WISE}\\
162.314540 & $-$53.318283 & 0.07 & 55379.3 & {\it WISE}\\
162.314283 & $-$53.318415 & 0.12 & 55567.3 & {\it WISE}\\
\enddata
\end{deluxetable}

\begin{deluxetable}{ll}
\tabletypesize{\scriptsize}
\tablewidth{0pt}
\tablecaption{Parallax, Proper Motion, and Photometry for
WISE~J104915.57$-$531906.1\label{tab:data}}
\tablehead{
\colhead{Parameter} & \colhead{Value} \\
}
\startdata
$\pi$ & 0.496$\pm$0.037$\arcsec$ \\
$\mu_{\alpha}$ cos $\delta$ & $-2.759\pm0.006\arcsec$~yr$^{-1}$ \\
$\mu_{\delta}$ & $+0.354\pm0.006\arcsec$~yr$^{-1}$ \\
DENIS $i$ & 14.94$\pm$0.03 \\
DENIS $J$ & 10.68$\pm$0.05 \\
DENIS $K_s$ & 8.87$\pm$0.08 \\
2MASS $J$ & 10.73$\pm$0.03 \\
2MASS $H$ & 9.56$\pm$0.03 \\
2MASS $K_s$ & 8.84$\pm$0.02 \\
$W1$ & 7.89$\pm$0.02 \\
$W2$ & 7.33$\pm$0.02 \\
$W3$ & 6.20$\pm$0.02 \\
$W4$ & 5.95$\pm$0.04 \\
\enddata
\tablecomments{$\pi$ and $\mu$ were derived from a fit to astrometry
from DSS, DENIS, 2MASS, and WISE (Section~\ref{sec:astro}).
The photometry is from the Third DENIS Release
(DENIS J104919.0$-$531910), the 2MASS Point Source
Catalog (2MASS J10491891$-$5319100), and the {\it WISE} All-Sky Source Catalog.}
\end{deluxetable}

\end{document}